# Design, fabrication, and spectral characterization of temperature-dependent liquid crystal-based metamaterial to tune dielectric metasurface resonances

*Golsa Mirbagheri[1], David T. Crouse[2]*

*Duke University Pratt School of Engineering[1]*
*Computer and Electrical Department of Clarkson University[2]*

## ABSTRACT

Tunable dielectric meta-surface nanostructures offer incredible performance in optical application due to their extraordinary tunability of the polarization and engineering the dispersion of light with low loss in infrared range. In this article, we designed and experimentally measured the tunability of all-dielectric subwavelength silicon nanoparticles with the help of the temperature-based refractive index of the liquid crystal in the telecom regime. The proposed structure composed of high dielectric nanodisk surrounded by nematic liquid crystal (NLC) is simulated with numerical software, assembled with pre-alignment material, and optically measured by Fourier-transform infrared (FTIR) spectroscopy. The simulated result is compatible with the practical measurements, shows that the tunability of 30nm is achieved. Electric and magnetic resonance modes of the high dielectric nanodisks are tailored in different rates by anisotropic temperature dependent NLC. The phase switching of anisotropic to isotropic nematic liquid crystal enables spectral tunning of the two modes of all dielectric metasurface and modifies the symmetry of the optical response of the metamaterial structure.

## 1. INTRODUCTION

**Magnetic and Electric Resonances Modes in Subwavelength High-dielectric Nanodisks**

In recent years, light tunability became one of the most important challenges in subwavelength optical devices, especially structures with lossy elements like nanometals which effect the optical application's functionality. Therefore, there is a need to investigate on alternative all-dielectric subwavelength optics instead of plasmonic nanostructure to compensate the strong losses of metals in sensing and subwavelength metamaterials applications. High index dielectric subwavelength nanodisks change the light scattering with the electric and magnetic resonance modes interference [1, 2, 3]. Despite the plasmonic metamaterials, all dielectric metamaterial nanostructure shows low loss at near-infrared spectral range when controlling the light waves. However, both metamaterials show strong resonances that can be tuned with respect to the aspect ratio of the metasurface layer, especially in case of high dielectric metasurfaces that support shifting the electric and magnetic modes excited from the circular movement of currents in the nanoparticles [4, 5, 6]. Compared to plasmonic nanostructure which resonant scattering excited only by electric resonances, low loss high dielectric nanoparticles have benefits of dynamically engineering the dispersion of light to near unity transmission with higher quality factor applicable in many applications including dynamically emission control systems and tunable optical filters [7, 8]. The individual silicon nanodisks embedded in low index medium show both electric and magnetic resonant modes which offers the destructive interference of resonant backward scattering and therefore, enhance the forward resonant light scattering when two resonant modes overlapping. By changing the geometric size of the high dielectric nanowires aspect ratio, the individual modes are tuned with respect to each other, this leads to manipulate the light at resonant or non-resonant of the desired spectra [1, 9, 10, 11].

---

[1] Send correspondence to Golsa Mirbagheri
E-mail: gm226@duke.edu

All dielectric high index nanoparticles show similar physics to metal-based nanostructures (e.g. split-ring resonator) in case of the magnetic Mie resonance resulted from the exciting the specific EM mode inside of the nanoparticles. According to Mie theory, the excitation of magnetic and electric modes happens when the effective wavelength of light inside the nanodisk is close to the nanoparticle's aspect ratio geometry, which shows electric field antiparallel to the nanoparticles with resonant magnetic field at center. This makes the magnetic wall boundary conditions that makes the electric resonance shift stronger than magnetic shift and therefore, this difference makes the two modes overlapped [1, 12, 13, 14]. Having both electric and magnetic resonance modes in photonic metamaterial have benefit of resonant scattering that supports the overlapping of two modes to increase the transmission of light and therefore, enhance forward scattering which is a significant advantage for many applicable devices like nanoantenna and sensing. This is in contrast with the regular scattering that shows the symmetric forward and backward scattering. Without special elements or substrate, backward scattering is supported by overlapping the two equal resonance modes by implementing the high index dielectric nanoparticles. The main mode of three-dimensional high dielectric nanoparticle is magnetic dipole which happens at the electric resonant frequency and therefore, these nanostructures can't shift the electric and magnetic resonant modes. However, a single high dielectric nano disk-shaped embedded into low dielectric material supports tailoring two separate modes with respect to their spectral position by changing the height/diameter aspect ratio and therefore, overlapping the lowest order resonance modes to suppress the backward scattering and improve transmission of the nanostructure [1, 15, 16, 17].

For easier comparison of the two modes of a single nanodisk excitation, we performed sweep analysis for the lossless filter with Ansys HFSS software. We considered the constant diameter of nanoparticle d=630nm and changed the lattice constant from 880nm to 980nm. The height of the nanodisk was considered as 220nm. The refractive index of silicon (and silicon substrate beneath) and $SiO_2$ were 3.5 and 1.45 respectively. The simulated result of measuring transmission of the normal light incident through the high dielectric nanocylinder-based structure is shown in Figure 1. By changing the lattice constant, the distance between crossing modes is changing [1, 18, 19, 20]. Figure 2 shows the electric field at electric resonant of the filter (1.52 µm) in addition to the magnetic resonant of the filter (1.44 µm) are illustrated in Figure 2.

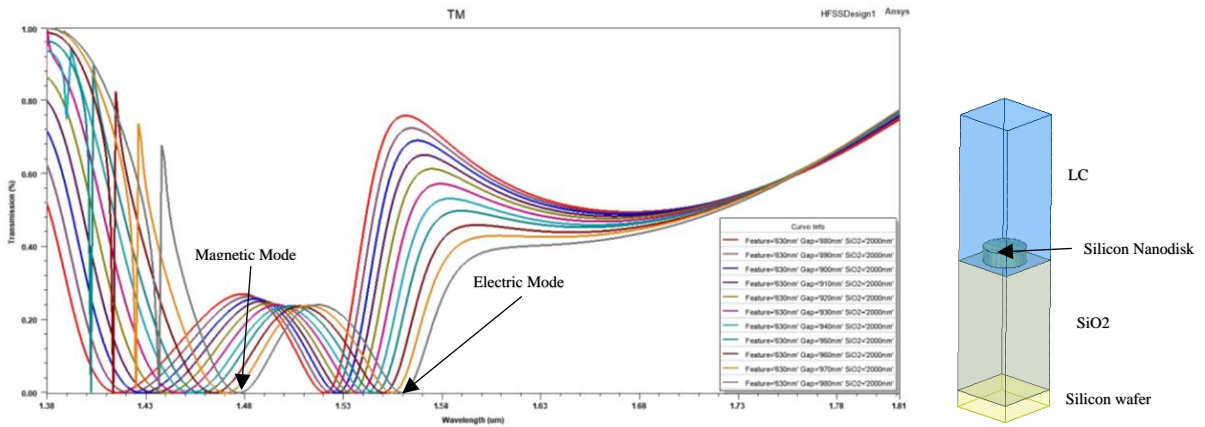

Figure 1. Transmission of the normal light incident through the lossless filter with Ansys HFSS. By changing the lattice constant from 880nm to 980nm, the distance between crossing modes is changing (Left). The unit cell of the high dielectric nano disk-based filter (Right).

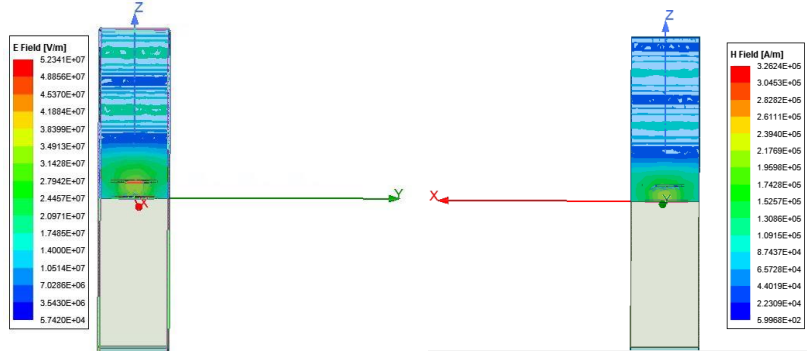

Figure 2. Electric field at 1.52 um (left) and Magnetic field at 1.44 um (right) of the excited modes
for high dielectric nanodisk

Different methods have been introduced to tailoring and shifting optical modes of lossless filters by applying external mechanical methods like temperature or voltage tuning. Furthermore, phase change materials like vanadium oxide can be used to change the phase of the metamaterials [21].And at last, temperature-based anisotropic liquid crystal, which we used in our work to dynamically control the resonance modes of the metamaterial structure [9, 10, 22].

## 2. NUMERICAL SIMULATION

The numerical calculation of the filter was performed by the Ansys HFSS electromagnetic simulation software using the Finite Element Method (FEM). The transmission of LC based filter is calculated by the master/slave boundaries that model the unit cell of periodic structures. The isotropic index in addition to extraordinary and ordinary indices of anisotropic phase of the 5CB LC are illustrated in Figure 3 at different temperatures [23].

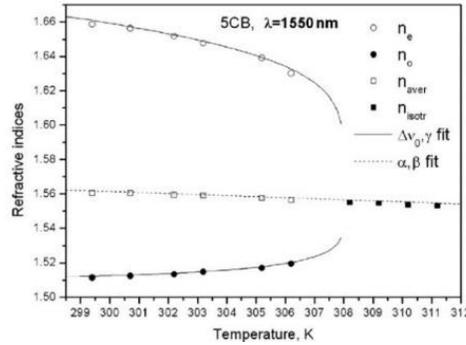

Figure 3. temperature dependent refractive indices of 5CB NLC at different temperatures [24]

The anisotropic refractive indices of LC were defined by Vuks model in Eq. 1, where $\bar{n}$ is average refractive index and $\Delta n$ is the birefringence equal to $n_e - n_o$ [25, 26, 27, 28]:

$$n_e = \bar{n} + \frac{2}{3}\Delta n$$

$$n_o = \bar{n} - \frac{1}{3}\Delta n \qquad (1)\ [25]$$

The metamaterial structure composed of 2μm anisotropic NLC is modeled and measured numerically with Ansys HFSS for both x and y polarization with the extraordinary index $n_e = n_{xx} = 1.69$ and ordinary index $n_o = n_{yy} = 1.53$ at room temperature 24°C, where the result is illustrated in red ($n_e$) and blue ($n_o$) lines in Figure 4. The refractive index of isotropic phase of the homogeneous liquid crystal is assumed to be as n = 1.58 at critical temperature 39°C, that is also illustrated in green line [29, 30, 31]. In the simulation the pre-alignment material is not included in the structure.

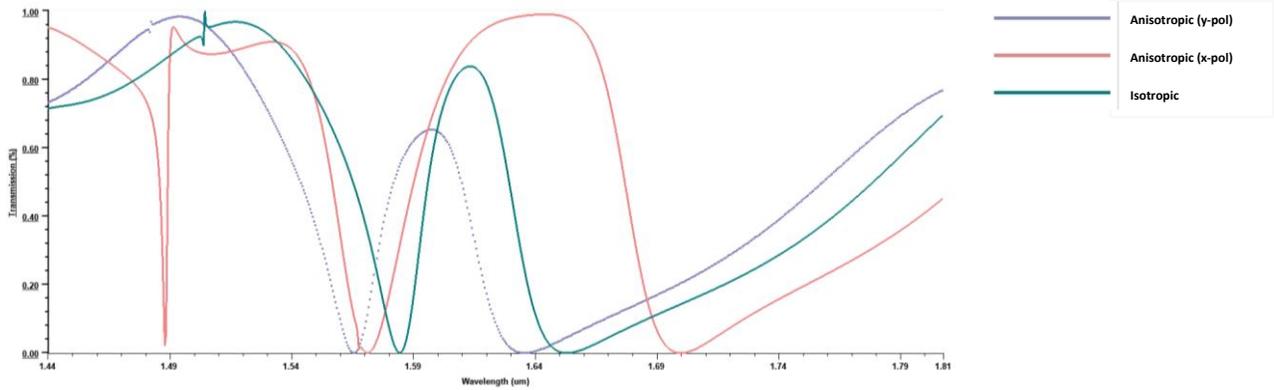

Figure 4. Transmission of Isotropic (green line) and Birefringence (peach line for x-polarization and blue dot-line for y-polarization) NLC modeled by Ansys HFSS

The height and diameter of the nanodisks are 220nm and 600nm respectively with lattice constant of 930nm. The top electrode is glass wafer with index n= 5.5, while the bottom electrode is the silicon wafer. The refractive index of silicon is considered n=3.5. The thickness of spacer dielectric silicon oxide layer is 2μm which was vapor deposited by PECVD method at 350°C. All materials of the simulated filter are simulation are supposed to be lossless (k=0). By infiltering the NLC into the filter, the magnetic and electric resonances are shifted significantly due to the higher refractive index of NLC to air. On the other hand, the temperature based NLC has the benefit of tuning the two different resonances instead of changing the aspect ratio of the high dielectric nano disks. In our work, the nematic liquid crystal (provided by Sigma Aldrich) has the height of 2 μm, which is injected between two electrodes by the capillary effect. The top electrode is glass which is covered by the pre-alignment material for planar alignment, shown in Figure 5.

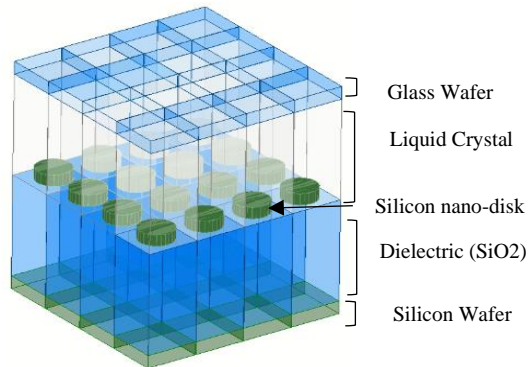

Figure 5. Subwavelength high dielectric metasurface-composed structure surrounded by NLC.

## 3. FABRICATION PROCESS

The 2μm silicon oxide on the bottom wafer was thermal deposited by the PECVD method with 20sccm SiH$_4$ and 2500sccm N$_2$O at pressure 1800Torr, temperature 350°C and power 140W. The amorphous silicon is deposited with the same method, with 25sccm SiH4 in 3mTorr and 10W at 300°C. The a-Si nanodisks were etched with 20sccm HBr at HF 30W and pressure 8mTorr, and then ached with oxygen gases. Figure 6 shows the top view of the etched a-Si with SEM.

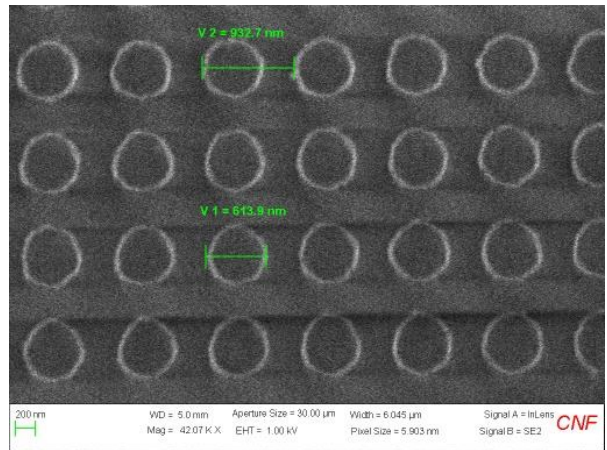
Figure 6. SEM top view of the etched a-Si

In this work, the 5CB nematic liquid crystal is provided by Sigma Aldrich. The NLC was injected on the meta-surface (between two electrodes) with capillary effect. The thickness of the LC was considered 5μm by the photoresist deposited and developed on the bottom electrode. Also, the spacer or $SiO_2$ spheres are recommended as alternatives in [22, 10]to define the thickness of the crystal. In the anisotropic phase of NLC, the orientation of crystal molecules is controlled by the electric filed between the two electrodes of the structure. To pre-align the NLC molecules, first a solution of the polyimide nylon 6 in 2,2,2 trichloroethanol by %1 mass concentration was prepared. Both Nylon 6 and trichloroethanol were provided from Sigma Aldrich. The pre-alignment solution was steered on agitator for a week. Then it was spin-coated on the top electrode and the baked at 185C for 70minutes. The baked solution was rubbed with the %100 cotton tissue (or velvet cloth) 10 times in planar direction (in our project in Y direction).

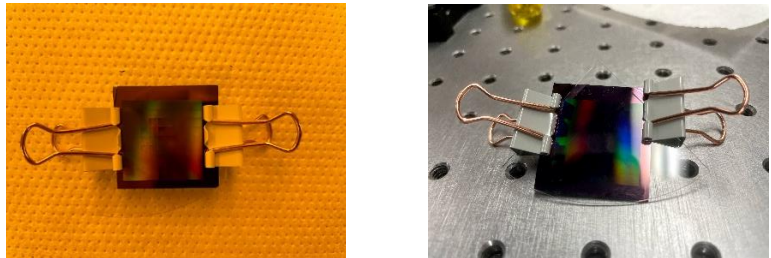
Figure 7. The liquid crystal injected between two wafers hold with two clips

Figure 7 shows the final fabricated cell composed of the NLC between the top electrode with spin-coated pre-alignment solution and bottom grating wafer. Rubbing the cotton tissue on the solution in the parallel direction to surface of the wafer creates the microscopic grooves that guides the crystal molecule fall into the vicinity holes along to the rubbing direction, as shown in Figure 8 [32, 33, 34].

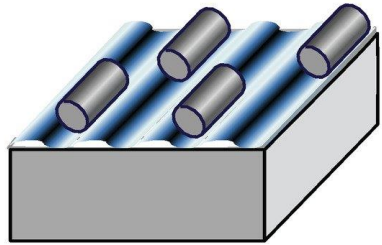
Figure 8. LC molecules oriented in the direction of the pre-alignment layer grooves (blue) [32].

**Related fabrication work for LC**

Different methods were investigated to assemble the liquid crystal between the two wafers. In [35], the glass wafers were coated and aligned with two methods to define the LC direction. In the homeotropic alignment, the LC is aligned in the orthogonal direction to the glass plate (Figure 9, left). In the planar alignment, the LC is aligned parallel to the glass wafer (Figure 9, right). The homeotropic alignment were performed with polyimide (SE-4811, Nissan Chemical) and the planar alignment was performed with a solution of 1 wt% Nylon 6,6 and 2,2,2 trichloroethanol that was spin coated on the wafer, then baked at 180°C for 4 hours and finally rubbed by a velvet wipe to make the tiny tilt (direction) at the surface in [35]. The LC was injected by capillary effect and the thickness of LC was defined by the spherical spacers.

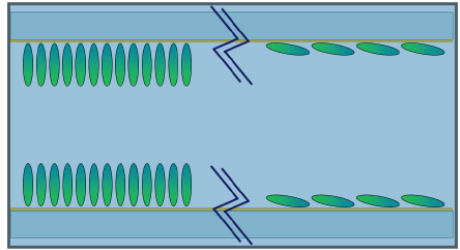

Figure 9. Homotropic alignment (left) and planar alignment (right) of the liquid crystal on the substrate [35]

In [36], polyvinyl alcohol (PVA) is used as pre-alignment layer, spin coated and brushed on the ITO layer of the filter to make the desired LC alignment. Since PVA scratches the metasurface, just one electrode was coated with the solution. The LC anisotropic phase is changed to isotropic by tuning the external sinusoidal AC voltage through ITO layer, where the PVA layer aligned perpendicular to the mirror plane of the split ring resonator (SRR) metasurface on the bottom electrode. The spacers were used to determine the Licristal E7 LC thickness [36, 37]. In [38], polymidies such as nylon-6, nylon-6,6 and nylon-6,10 are introduced as good orienting layers for nematic liquid crystals. The solubility of nylon-6 in trifluoroethanol was investigated, shows less viscous than trichloroethanol solution. This makes the solution of nylon6 and trifluoroethanol an excellent homogenous alignment to be rubbed on the surface unidirectionally. In [39] non-stoichiometric evaporated oxide such as $SiO_x$ is intruded as alignment layer which shows voltage shift response in the corresponding structure. The $SiO_2$ isolating material in [39] was deposited on the ITO layer on one wafer for the homogeneous planar alignment, while the other glass electrode was coated by fluorinated polystyrene block copolymer. In [40] the thickness of E7 LC injected between two electrodes determined by the round plastic balls. The planar alignment performed with the spin coated nylon6 solution which was rubbed unidirectionally on the proposed wafer. The injected NLC changed the phase to isotropic state at 58 °C.

In [41, 42], two electrodes were covered with planar aligned material in orthogonal direction and filled with a thick NLC and an extra photopolymer layer to measure the different directions of LC alignment. In [43], the electrodes were spin coated with organic polymer materials like polyimide (beside rubbed process), un−rubbed nylon 6 or inorganic material like oblique evaporated $SiO_X$ or $SiO_2$ which are less effected by the UV radiation or hot environment. In [44], the polarization direction of reflective metamaterial twisted nematic liquid crystal cell (RMNLC) is affected by the direction of twisted nematic crystal. The twisted nematic crystal injected between polyimide coated ITO glass and rubbed PVA coated quartz wafer behaves as a polarization rotator. In [45], chiral liquid crystal mixtures with chiral dopant 3.7 wt% is injected between two ITO-covered glasses using the capillary effect. The cholesteric liquid-crystal (CLC) is placed between two rubbed nylon electrodes in the planar direction, then exposed to 365 nm UV light to be polymerized. By changing the concentration of the NLC and UV dose, CLC characteristic can be controlled.

## 4. MEASUREMENT

For measuring the transmittance of incident light of the sample, the FTIR spectrometer was applied [46]. The mercuric cadmium telluride (MCT) detector and $CaF_2$ beam splitter in FTIR measured the optical properties of the LC-based metamaterial filter. The temperature of the sample was tuned from room temperature to the temperature higher than critical level by help of the temperature controller tool which is attached to a sample holder and put into the reflector cell inside of the FTIR, as shown in Figure 10. The heat was transferred from the back silicon wafer which was on the sample holder of the temperature controller.

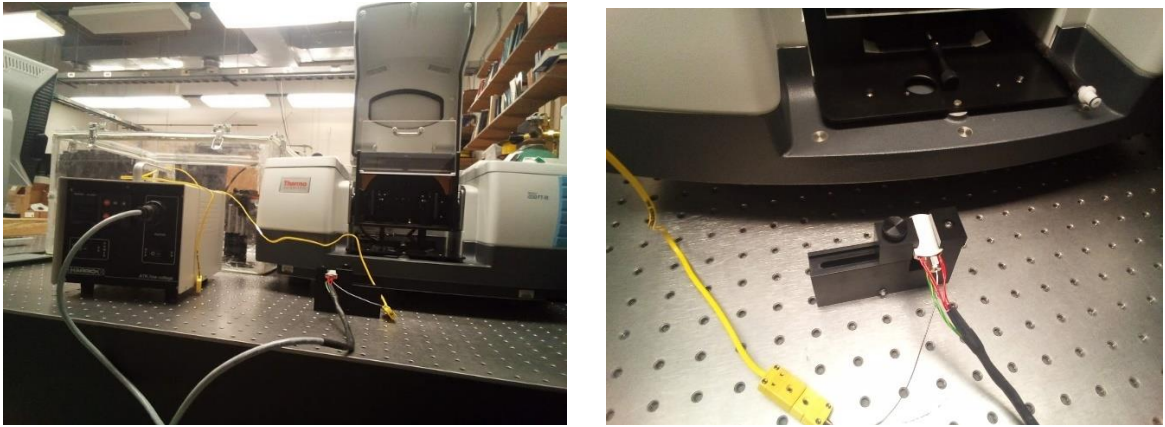

Figure 10. Schematic of the Temperature Controller (Left) with sample holder (Right)

A wire-grid polarizer was inserted in input port of the reflector cell to define the incident light polarization. The incident light was guided through the bottom silicon wafer in such a way that light first reached to the silicon nanodisks and excite them before reaching to the LC and therefore, was independent to the orientation of the NLC. For the background spectra, the other filter area without metasurfaces was used to collect the spectra data.

To calculate the transmittance of the assembled filter, the FTIR spectroscopy was implemented. For the anisotropic phase of NLC in room temperature $24°C$, we setup up polarizer before the sample, in such a way that TM polarized entrance light was transmitted through the anisotropic LC based filter. The polarization of the incidence light is parallel to the anisotropy axis of the aligned LC on the electrode. The extraordinary light impinged on the LC first before hitting meta surface, then focused on the LC sample. The reflected light then detected by the MCT detector. The temperature of the sample is adjusted by the temperature controller.

In Figure 11, the shifts of the magnetic dipole resonance and the electric dipole resonance related to the temperature tuning are represented for the simulated structure with HFSS and the experimental, measurement with FTIR. In the simulated structure at the right, the magnetic resonant of the anisotropic liquid crystal, which is aligned parallel with the electric field, is at the 1.57 µm, while the electric dipole resonant is placed at the 1.70 µm. By changing the anisotropic material to isotropic in HFSS, we see a considerable change in the magnetic dipole resonance (MR) and electric dipole resonance (ER) wavelength. However, the 50 nm ER shift is more extensive than MR shift. The assemble structure is also measured with Fourier spectroscopy method, plotted with MATLAB in Figure 11 at right. We used separate reference spectrum for each measurement using the mirror as the reference. The lower transmission is due to the rubbing process which scratch the wafer. For this problem, the AtA-2 alignment material in addition to DUV step are recommended to avoiding scratching the wafer in [10, 47, 14, 48, 49]. Furthermore, the deviation of the result can be explained by the unfitted alignment of the fabricated filter.

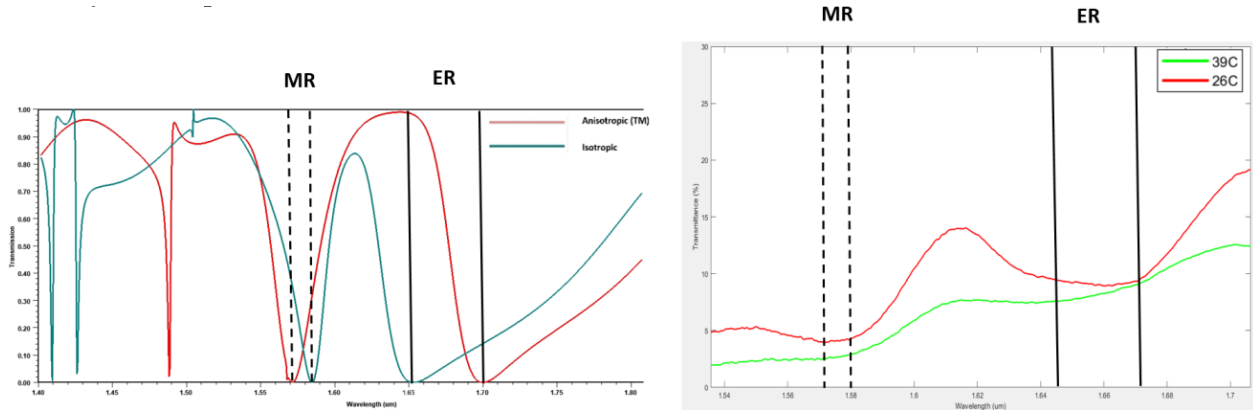

Figure 11. MR and the ER shifts due to temperature tuning for simulated structure
with HFSS (left) and experimentally measurement with FTIR (right)

However, we still clearly observe the MR and ER shifts of the silicon nanoparticles after implying the different temperature. Since the isotropic phase of the 5CB nematic LC appear at the temperature 35 ℃, we can measure the phase change at any temperature below or above this critical level. At temperature 26 ℃, near room temperature, the anisotropic NLC has two MR and ER dipole mode in 1.57 µm and 1.67 µm. By increasing the temperature to 39 ℃, the spectral positions of the dipole resonances move closer to each other, this indicates that the molecules of the NLC reoriented into the grooves of the pre-alignment material and shape a uniform structure, shows the shifts of the electric and magnetic modes of the NLC are 30 nm and 10nm. The adjustment of magnetic and electric resonance of the nano cylinders can be applied in so many dynamic applications, including the tunable imaging systems.

## 5. CONCLUSION

In this article, we designed all dielectric LC-based structure to tune and experimentally measure the electric and magnetic resonance of the high dielectric nano particles with a maximum tuning range of 30 nm. By applying different temperatures from room temperature to state change level (35 ℃), the NLC molecules goes under two different phases. With the state transition of injected NLC from anisotropic to isotropic dynamically, the symmetry of the structure breaks which leads to the displacement of the two dipole resonances in silicon disks dynamically without changing the geometric features of the element. The experimental result has agreement with the proposed simulated model that was simulated with anisotropic uniaxial refractive index tensor related for of the NLC molecules. Changing the magnetic and electric resonance position in respect to each other can be performed faster with applying external voltage transition as well which is considered in many tunable silicon metasurfaces devices in near IR. The dynamically shifting the dipole resonance with the help of LC can be applicable in broad range of switchable all dielectric metamaterial structures.

## 6. ACKNOWLEDGEMENT

This work was performed in part at the Cornell NanoScale Facility, a member of the National Nanotechnology Coordinated Infrastructure (NNCI), which is supported by the National Science Foundation (Grant NNCI-2025233). We would like to thank the NSF Industry/University Cooperative Research Center for Metamaterials to support this project.